\documentclass{article}
\usepackage[cp1251]{inputenc}
\usepackage[english]{babel}
\usepackage{amsmath,amsthm}
\usepackage{amsfonts}
\usepackage{graphicx}
\usepackage{url}
\usepackage{hyperref}

\title{Music Recommendation System for Million Song Dataset Challenge}

\author{Nikolay Glazyrin \\ Ural Federal University, Ekaterinburg \\ {\tt
nglazyrin@gmail.com}}

\begin{document}

\maketitle

\begin{abstract}
In this paper a system that took 8th place in Million Song Dataset
challenge is described. Given full listening history for 1 million of users and
half of listening history for 110000 users participatints should predict the
missing half. The system proposed here uses memory-based collaborative filtering
approach and user-based similarity. MAP@500 score of 0.15037 was achieved.
\end{abstract}

\section{Introduction}

The goal of the Million Song Dataset Challenge \cite{Mcfee2012} was to restore
the full listening history of 110000 users given half of their listening history
and the full history for 1019318 other users. This full listening history
contains 48373586 unique triplets $user\_id, track\_id, play\_count$. All the
tracks belong to the Million Song Dataset, a freely-available collection of
audio features and metadata for a million contemporary popular music tracks
\cite{Bertin-Mahieux2011}. Additional information about tracks, such as author,
song title, year, tags, song text, was provided by SecondHandSongs
\footnote{\url{http://www.secondhandsongs.com/}}, musiXmatch
\footnote{\url{http://musixmatch.com/}}, last.fm
\footnote{\url{http://www.last.fm/}}, The Echo Nest
\footnote{\url{http://the.echonest.com/}}.

\subsection{Evaluation}

Mean Average Precision truncated at 500 was used to evaluate
systems' recommendations \cite{Mcfee2012}. Let $M$ be the user-track matrix,
where $M_{u,t}=1$ if and only if user $u$ listened to the track $t$, and
$M_{u,t} = 0$ otherwise. Let $y$ be a ranking over items, $y(j) = i$ means that
item $i$ is ranked at the position $j$. For each $k \leq 500$ \textit{precision
at k} $P_k(u)$ can be defined as the number of correct recommendations within
first $k$ recommendations divided by $k$:
$$
P_k(u,y) = \frac{1}{k} \sum \limits_{j=1}^k M_{u,y(j)}
$$

Then \textit{average precision} can be defined as
$$
AP(u,y) = \frac{1}{n_u} \sum\limits_{k=1}^\tau P_k(u,y) \cdot M_{u,y(k)}
$$
Here $n_u$ is a minimum of 500 and the total number of tracks recommended to the
user $u$. Then by averaging $AP(u,y)$ over all users the \textit{mean average
precision} can be obtained:
$$mAP = \frac{1}{n} \sum\limits_u AP(u,y)$$
Here $n$ is the total number of the users. $N=110000$ during the evaluation.

\section{System description}

For the ease of comparison we will follow here the notation used by F. Aiolli in
\cite{Aiolli2012}. Let $\mathcal{U}$ be the set of all users, and $\mathcal{I}$
--- the set of all tracks. In the proposed system the known half of listening
history for test users was also used to generate recommendations, so
$|\mathcal{U}| = n = 1129318$, $|\mathcal{I}| = m = 384546$. $R=\{r_{ui}\} \in
\mathbb{R}^{n \times m}$ represents how much user $u$ likes track $i$. But here
we assume that $r_{ui} \in [0, 1]$.

The proposed system implements user-based recommendation. The scoring function
for this recommendation type is computed by
$$
h_{ui}^U = \sum\limits_{v \in \mathcal{U}} f(w_{uv}) r_{vi}
$$
Here $w_{uv}$ denotes the similarity between users $u$ and $v$. We use the
identity function instead of $f(w)$, so the scoring function can be written as
$$
h_{ui}^U = \sum\limits_{v \in \mathcal{U}} w_{uv} r_{vi}
$$

Now the goal is to choose $w_{uv}$ and $r_{vi}$ to generate better
recommendations to the user $u$. The recommended tracks can be sorted by the
value of $h_{ui}^U$ in decreasing order. Then the resulting list is truncated at
500 items or supplemented with dummy track ids (1, 2, \ldots). It was not a
frequent case when the list of recommendation contained less than 500 items,
roughly 1 time per 1000 users. So we do not expect this simple supplement to
reduce recommendation quality considerably.

\subsection{Analogies to textual information retrieval}

Some analogies can be drawn between songs recommendation and textual search. The
listening history for a user $u$ can be considered as a \textit{document}, where
each single listened track is a \textit{term}, and the number of listenings for
this track is the number of occurences of this term in the document. All the
documents make a \textit{collection}, which, in our case, corresponds to the
whole listening history of all users.

In the information retrieval field \textit{term frequency} $tf_{t,d}$ is the
number of occurences of a term $t$ in a document $d$. It corresponds to the
number of listenings of a track $t$ by a user $u$. \textit{Document frequency}
$df_t$ is the number of documents in the collection that contain a term $t$ ---
the number of users who listened to the track $t$ in our case. Then the
\textit{inverse document frequency} $idf_t$ is defined as
$$
idf_t = log \frac{n}{df_t}
$$
Here $n$ is the total number of documents (users). More details can be found in
\cite{Manning:2008:IIR:1394399}. We will use the term \textit{idf} here with
regard to tracks instead of terms.

It can be seen from the definition of idf that its value is high for rare
tracks, which were listened by only a few users. Intuitively, if two users
listened to the same track and nobody else did it, these users are probably
similar. For the popular tracks the value of idf will be small. These tracks
should not contribute much to users similarity.

\subsection{User similarity and track ranks}

Bearing in mind these conclusions, we can define the similarity $w_{uv}$ between
users $u$ and $v$ as the sum of idf of the tracks that they have listened both.
Let $\overline{w}_u$ be the value of similarity of the most similar user to a
user $u$. All the users whose similarity value is less than $s \cdot
\overline{w}_u$ are discarded. The value of $s=0.4$ was chosen here.

The value $r_{vi}$ of how much a user $v$ likes a track $i$ is chosen as
$\frac{1}{c_v}$ if user $v$ listened to a track $i$ and 0 otherwise. Here $c_v$
is the total number of listenings for a user $v$ --- the sum of all play counts
from his listening history. So all the tracks are ranked equally regardless of
their play counts. The resulting scoring function can be written as
$$
h_{ui}^U = \sum\limits_{v \in \mathcal{U}_s}
\left(\sum\limits_{\substack{i:M(u,i) > 0 \\ M(v,i) > 0}} idf_i \right) \frac{1}{c_v}
$$
Here $\mathcal{U}_s$ is the set of users whose similarity (the sum in brackets)
is greater than $s \cdot \overline{w}_u$.

\subsection{Results}

This algorithm achieved the final score of $0.15037$ on the whole test set. This
corresponds to 8th place in the leaderbord. The process of generation of
recommendations for 110000 users took 3 hours on a PC with Intel Core i5 @ 2.8
GHz processor running Windows 7 x64 Professional and JDK 1.6.0\_29 64-bit. The
program is single-threaded. The memory consumption was constant, about 1 GB.

%
%
%
%

\bibliographystyle{plain}
\bibliography{admire,aiolli2012,Manning2008,millionsong}

\end{document}